\documentclass{JHEP3}
\usepackage{graphicx}
\usepackage{latexsym}
\def\be{\begin{equation}}
\def\ee{\end{equation}}
\def\bea{\begin{eqnarray}}
\def\eea{\end{eqnarray}}
\def\l{\left}
\def\r{\right}

\title{High-Redshift Cosmography}
\author{Vincenzo Vitagliano${}^{1,2}$, Jun-Qing Xia${}^1$, Stefano Liberati${}^{1,2}$, Matteo Viel${}^{2,3}$ \\
${}^1$SISSA, Via Beirut 2-4, 34151 Trieste, Italy\\
${}^2$INFN sez. Trieste, Via Valerio 2, 34127 Trieste, Italy\\
${}^3$INAF-Osservatorio Astronomico di Trieste, Via G.B. Tiepolo 11, I-34131 Trieste, Italy \\
Email: \email{vitaglia@sissa.it}, \email{xia@sissa.it}, \email{liberati@sissa.it}, \email{viel@oats.inaf.it}}

\abstract{We constrain the parameters describing the kinematical state
  of the universe using a cosmographic approach, which is fundamental
  in that it requires a very minimal set of assumptions (namely to
  specify a metric) and does not rely on the dynamical equations for
  gravity.  On the data side, we consider the most recent compilations
  of Supernovae and Gamma Ray Bursts catalogues. This allows to
  further extend the cosmographic fit up to $z=6.6$, i.e. up to
  redshift for which one could start to resolve the low $z$ degeneracy
  among competing cosmological models.  In order to reliably control
  the cosmographic approach at high redshifts, we adopt the expansion
  in the improved parameter $y=z/(1+z)$. This series has the great advantage to hold also
  for $z>1$ and hence it is the appropriate tool for handling data
  including non-nearby distance indicators.  We find that Gamma Ray
  Bursts, probing higher redshifts than Supernovae, have constraining
  power and do require (and statistically allow) a cosmographic
  expansion at higher order than Supernovae alone. Exploiting the set of data from Union and GRBs catalogues, we show (for the first time in a purely cosmographic approach parametrized by deceleration $q_0$, jerk $j_0$, snap $s_0$) a definitively negative deceleration parameter $q_0$ up to the $3 \sigma$ confidence level. We present also
  forecasts for realistic data sets that are likely to be obtained in
  the next few years.}

\begin{document}

\section{Introduction}

A huge amount of cosmological data sets with increasing reliability
has been collected during the last decade, providing new insights on
the dynamical state of our universe. The interpretation of the Hubble
diagram for Type Ia Supernovae (SNeIa) and the observations of
polarization and anisotropies in the power spectrum of the Cosmic
Microwave Background (CMB) showed that the universe is undergoing an
accelerated phase of expansion. Furthermore, the availability of new
cosmological probes, such as high redshift SNe and Gamma-Ray Bursts
(GRBs), enables to improve the knowledge of the cosmological
parameters and allows to distinguish among different models: any new
test should be a challenge for all the attempts providing an
explanation of the current cosmic acceleration \cite{exp}. On the other hand, the
recent development of a model independent approach (the cosmographic one
\cite{chiba, vc_taylor, vc, Rapetti:2006fv}) gained increasing interest for catching as much information
as possible directly from observations, retaining the minimal priors
of isotropy and homogeneity and leaving aside any other
assumption. Let us stress that the idea supporting cosmography is
rather clean: here we are not providing any new model for Dark Energy,
but just a deeper, even if simpler, analysis of the cosmological data
sets with the aim to give a fiducial frame against which any
theoretical proposal should be tested. The only ingredient taken into
account \textit{a priori} is the FLRW line element obtained from
kinematical requirements \be
ds^2=-c^2dt^2+a^2(t)\l[\frac{dr^2}{1-kr^2}+r^2d\Omega^2\r]\,; \ee
exploiting this metric, it is possible to express the luminosity
distance $d_L$ as a power series in the redshift parameter $z$, the
coefficients of the expansion being functions of the scale factor
$a(t)$ and its higher order derivatives.

In this paper we will discuss the use of luminosity distance
indicators in the high redshift universe in order to constrain the
values of fundamental cosmological parameters that describe the model
we are interested in.

Even though the prominent role of
SNe (in the high-$z$ version, too) in doing the job is well-known, the
potentiality of GRBs as cosmological standard candles has been
recently explored as a possible proposal to increase the number of
high redshift distance ladders. Data coming from observations of both
SNe and GRBs are used to fit directly the expression for the
luminosity distance $d_L$.

Following \cite{wei}, $d_L$ can be defined from the relation between
the apparent luminosity $l$ of an object and its absolute luminosity
$L$ \bea l=\frac{L}{4 \pi r^2_1 a^2(t_0)(1+z)^2}= \frac{L}{4\pi
d_L^2}\,, \eea where $r_1$ is the comoving radius of the light
source emitting at time $t_1$, $t_0$ is the later time an observer
in $r=0$ is catching the photons, and redshift $z$ is, as usual,
defined as $1+z=a(t_0)/a(t_1)$. The radial coordinate $r_1$ in a
FLRW universe can be written for small distances as \cite{v1} \be
r_1=\int_{t_1}^{t_0}\frac{c}{a(t)}dt-\frac{k}{3!}\l[\int_{t_1}^{t_0}\frac{c}{a(t)}dt\r]^3+\mathcal{O}(5)\,,
\ee with $k=-1, 0, +1$ respectively for hyperspherical, Euclidean or
spherical universe. In such a way, it is possible to recover the
expansion of $d_L$ for small $z$ \bea
d_L(z)=&&cH_0^{-1}\l\{z+\frac{1}{2}(1-q_0)z^2-\frac{1}{6}\l(1-q_0-3q_0^2+j_0+
\frac{kc^2}{H_0^2a^2(t_0)}\r)z^3+\r.\nonumber \\ && \nonumber \\
&&\hspace{-1cm}\l.+\frac{1}{24}\l[2-2q_0-15q_0^2-15q_0^3+5j_0+10q_0j_0+
s_0+\frac{2kc^2(1+3q_0)}{H_0^2a^2(t_0)}\r]z^4+...\r\}\,,
\eea
having defined the cosmographic parameters as:
\bea
H_0&\equiv&\l.\frac{1}{a(t)}\frac{da(t)}{dt}\right|_{t=t_0}\equiv\l.\frac{\dot{a}(t)}{a(t)}\r|_{t=t_0}~,\label{eq:H0}\\
q_0&\equiv&-\l.\frac{1}{H^2}\frac{1}{a(t)}\frac{d^2a(t)}{dt^2}\right|_{t=t_0}\equiv\l.-\frac{1}{H^2}\frac{\ddot{a}(t)}{a(t)}\r|_{t=t_0}~,\label{eq:q0}\\
j_0&\equiv&\l.\frac{1}{H^3}\frac{1}{a(t)}\frac{d^3a(t)}{dt^3}\right|_{t=t_0}\equiv\l.\frac{1}{H^3}\frac{a^{(3)}(t)}{a(t)}\r|_{t=t_0}~,\label{eq:j0}\\
s_0&\equiv&\l.\frac{1}{H^4}\frac{1}{a(t)}\frac{d^4a(t)}{dt^4}\right|_{t=t_0}\equiv\l.\frac{1}{H^4}\frac{a^{(4)}(t)}{a(t)}\r|_{t=t_0}~.\label{eq:s0}
\eea

It is important to stress the {existence} of other  definitions for distance indicators
\cite{vc} rather than the more often used $d_L$: depending on which physical
quantity one is measuring, it could be more convenient to extract from
data one of these instead of the luminosity distance. These different
quantities admit different expressions of the Taylor expansion in the
redshift, such that it could be more natural to estimate cosmographic
parameters, whose expression instead does not depend on the analytic
expansion, in one of these particular frameworks. {This ambiguity could lead to a misunderstanding even about the proper
definition of distance one should retain}. From now on we will refer
only to luminosity distance since it is the most direct choice in the
case of measures of distance for SNe and GRBs.

In recent works \cite{lucariello,wang-dai}, a first attempt to fit luminosity distance by data using more distant
objects as ``standard candles'' has been performed. However, the ill-behaviour at high
$z$ of the redshift expansion used there is known to strongly affect the results.  
We are no longer going to use the standard relation linking the luminosity distance to the ordinary-defined redshift. As already
pointed out in \cite{vc_taylor}, the lack of validity of the Taylor-expanded expression for $d_L$ could be settled down
approximately at $z\sim1$. In order to avoid problems with the
convergence of the series for the highest redshift objects as well as
to control properly the approximation induced by truncations of the
expansions, it is useful to recast $d_L$ as a function of an improved
parameter $y=z/(1+z)$~\cite{vc_taylor,CPL}. In such a way, being $z\in(0,\infty)$ mapped into
$y\in(0,1)$, we will be in principle able to retrieve the right
behaviour for series convergence at any distance. The introduction of
this new redshift variable will not affect the definition of
cosmographic parameters, while the luminosity distance at fourth order
in the $y$-parameter becomes:
\bea
d_L(y)=&&\frac{c}{H_0}\left\{y-\frac{1}{2}(q_0-3)y^2+\frac{1}{6}\left[12-5q_0+3q^2_0-(j_0+\Omega_0)\right]y^3+\frac{1}{24}\left[60-7j_0-\right.\right.\nonumber
  \\ &&\left.\left.-10\Omega_0-32q_0+10q_0j_0+6q_0\Omega_0+21q^2_0-15q^3_0+s_0\right]y^4+\mathcal{O}(y^5)\right\}~,
\eea
where $\Omega_0=1+kc^2/H_0^2a^2(t_0)$ is the total energy density.
For the flat universe, $\Omega_0=1$. On the other hand, luminosity
distance can also be expressed as ``logarithmic Hubble relations":
\bea \ln{\left[\frac{d_L(y)}{y~\rm
      Mpc}\right]}=&&\frac{\ln{10}}{5}\left[\mu(y)-25\right]-\ln{y}=\ln{\left[\frac{c}{H_0}\right]}
-\frac{1}{2}(q_0-3)y+\frac{1}{24}\left[21-4(j_0+\Omega_0)+\right.\nonumber
  \\ &&\left.+q_0(9q_0-2)\right]y^2+\frac{1}{24}\left[15+4\Omega_0(q_0-1)+j_0(8q_0-1)-\right.\nonumber\\ &&\left.-5q_0+2q^2_0-10q^3_0+s_0\right]y^3+\mathcal{O}(y^4)~;\label{eq:y-exp}
\eea
and then, the distance modulus (in which are expressed current data) is given by the expression:
\bea
\mu(y)=&&25+\frac{5}{\ln{10}}\left\{\ln{\left[\frac{c}{H_0}\right]}+\ln{y}
-\frac{1}{2}(q_0-3)y+\frac{1}{24}\left[21-4(j_0+\Omega_0)+q_0(9q_0-2)\right]y^2+\right.\nonumber
\\ &&\hspace{-.5cm}+\left.\frac{1}{24}\left[15+4\Omega_0(q_0-1)+j_0(8q_0-1)-5q_0+2q^2_0-10q^3_0+s_0\right]y^3+\mathcal{O}(y^4)\right\}~.
\label{dm}
\eea

{It is clear that the higher the order reached in the redshift
  expansion of $d_L$, the better the data fitting will be (more free
  parameters).  However, for a given data set there will be an upper
  bound on the order of the expansion which is statistically
  significant in fitting the data.  In the following sections we
  shall fit the data with two different truncations of expansion
  (\ref{eq:y-exp}) without and with the third order term ($y^3$). For
  the sake of conciseness we shall respectively label them as
  Cosmography I and Cosmography II. While we shall see that the higher
  redshift data make the latter truncation the most statistically
  significant among the two, we shall show in the end that no
  improvement is at the moment achievable by extending the truncation
  to fourth order in $y$.}

\section{Supernovae and GRB Datasets}

The SNeIa distance moduli provide the luminosity distance as a function
of redshift. In this paper we will use two different SNIa datasets:
one is the SuperNova Legacy Survey (SNLS) data over a redshift range
$z=0-1$ which consists of 115 data points \cite{snls}, and the other
one is the recently released Union compilation (307 data points, with systematic errors included) from
the Supernova Cosmology project \cite{Kowalski:2008ez}, which includes
the recent samples of SNeIa from the SNLS and ESSENCE survey, as well
as some older data sets, and span the redshift range
$0\lesssim{z}\lesssim1.55$. We do not include the data from CfA3 as their reliability is
controversial (see e.g.~\cite{cfa} and \cite{cfatension} for opposite points of view). In any case, being extremely low redshift
($z<0.1$), their removal does not strongly affect our analysis since those redshifts are already very well covered  
by the other catalogues.


In addition, we consider GRBs that can potentially be used to measure
the luminosity distance out to higher redshift than SNeIa.  GRBs are
not standard candles since their isotropic equivalent energetics and
luminosities span 3-4 orders of magnitude. However, similarly to SNeIa
it has been proposed to use correlations between various properties of
the prompt emission and also of the afterglow emission to standardize
GRB energetics (e.g. \cite{grb1}). Recently, several empirical
correlations between GRB observables were reported, and these findings
have triggered intensive studies on the possibility of using GRBs as
cosmological ``standard'' candles (Ref.\,\cite{lucariello},
\textit{e.g.}, finds the cosmographic parameters from luminosity
distance data obtained exploiting different relations among GRBs
quantities.).

For example, the GRB isotropic luminosities can be
computed as a function of redshift using the bolometric fluence and
the two exhibit strong correlations with the rest frame peak
energy. However, due to the lack of low-redshift long GRB data to
calibrate these relations, in a cosmology-independent way, the
parameters of the reported correlations are given assuming an input
cosmology and obviously depend on the same cosmological parameters
that we would like to constrain. Thus, applying such relations to
constrain cosmological parameters leads to biased results.

In Ref.~\cite{xiagrb} this ``circular problem'' is naturally eliminated
by marginalizing over the free parameters involved in the
correlations; in addition, some results show that these correlations
do not change significantly for a wide range of cosmological
parameters \cite{Firmani,Schaefer:2006pa}. Therefore, in this paper we
use the 69 GRBs over a redshift range from $z=0.17-6.60$ presented in
Ref.~\cite{Schaefer:2006pa}, but we keep in mind the issues related to
the circular problem that are more extensively discussed in
Ref.~\cite{xiagrb} and also the fact that all the correlations used to
standardize GRBs have scatter and are poorly understood under the
physical point of view.  For a more extensive discussion and for a
full presentation of a GRB Hubble Diagram with the same sample that we
used we refer the reader to Section 4 of Ref.~\cite{Schaefer:2006pa}.

In the calculation of the likelihood from SNeIa and GRBs, which will be
presented in the next sections, we have marginalized for each class of object over the absolute
magnitude $M$  which is a nuisance parameter, as done in
Ref.~\cite{Goliath:2001af,SNMethod}:
\begin{equation}
\bar{\chi}^2=A-\frac{B^2}{C}+\ln\left(\frac{C}{2\pi}\right)~,
\end{equation}
where
\begin{eqnarray}
A=\sum_i\frac{(\mu^{\rm data}-\mu^{\rm th})^2}{\sigma^2_i}~,~~~
B=\sum_i\frac{\mu^{\rm data}-\mu^{\rm th}}{\sigma^2_i}~,~~~
C=\sum_i\frac{1}{\sigma^2_i}~.
\end{eqnarray}

In order to compute the likelihood, we use a Monte Carlo Markov Chain technique as it is
  usually done to explore efficiently a multi-dimensional
  parameter space in a Bayesian framework (this procedure has been
  used only when the number of parameters becomes larger than three,
  otherwise direct calculation of the likelihood has been performed). For each Monte Carlo Markov Chain calculation, we run four independent chains that consist  of about $300.000\div500.000$ chain elements each. We test the convergence of the chains by Gelman and Rubin criterion \cite{gel} finding $R - 1$ of order $0.01$, which is more conservative than the often used and recommended value $R - 1 < 0.1$ for standard cosmological calculations.

Note that the error budget of the data we used also includes the
uncertainty in the redshift determination and the contribution of the
host-galaxy peculiar velocity for the SNe as done in \cite{Kowalski:2008ez} and
\cite{lampeitl}. As for the GRBs we refer the reader to \cite{Schaefer:2006pa} for
a discussion on possible systematic errors and their effect on the
luminosity distance.

\section{Statistical Analysis}

In this section we present our main results on the constraints for the
cosmographic expansion from the current observational data sets.

\subsection{Cosmography I}\label{subsec:C1}
\begin{figure}[htbp]
\begin{center}
\includegraphics[scale=0.36]{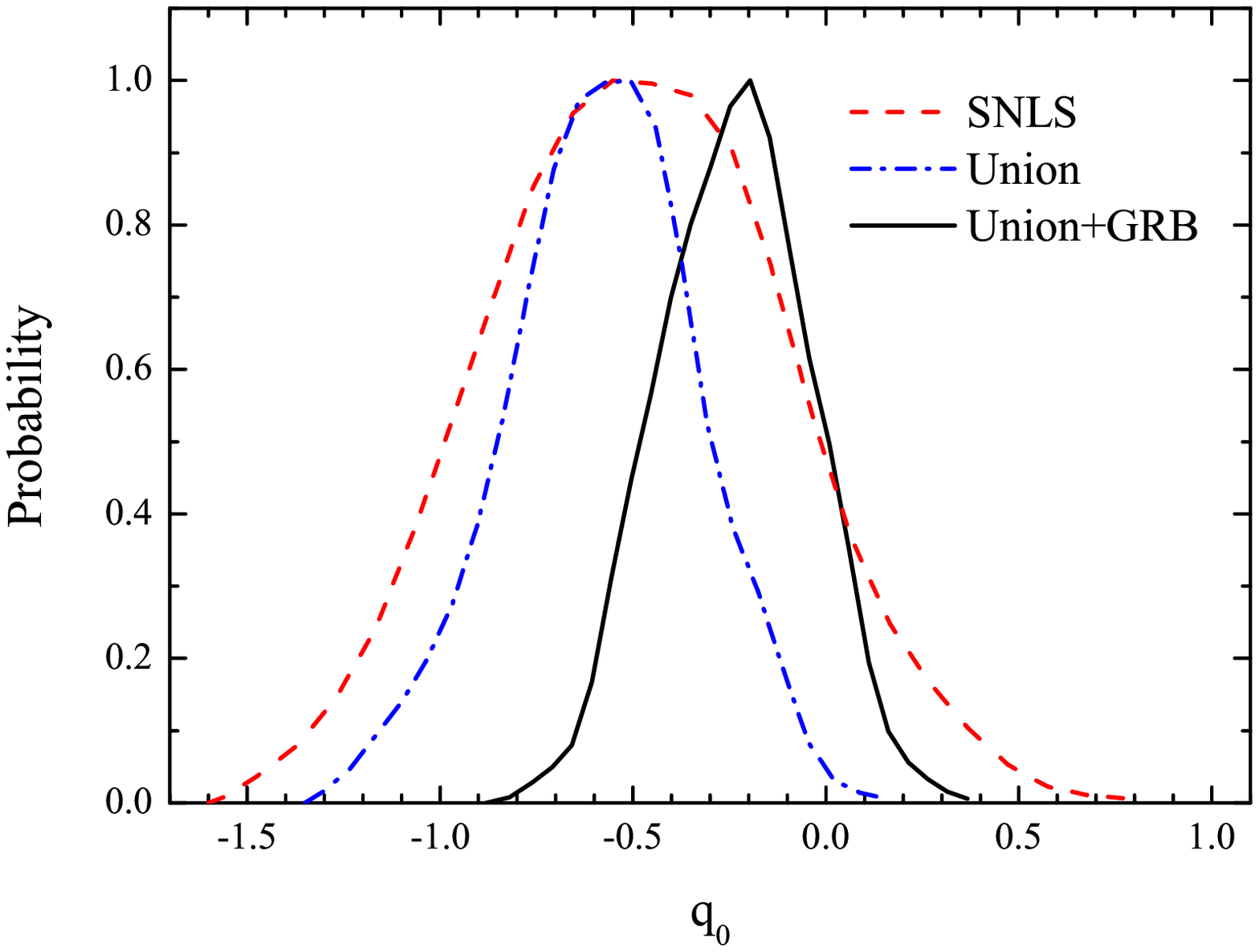}
\includegraphics[scale=0.36]{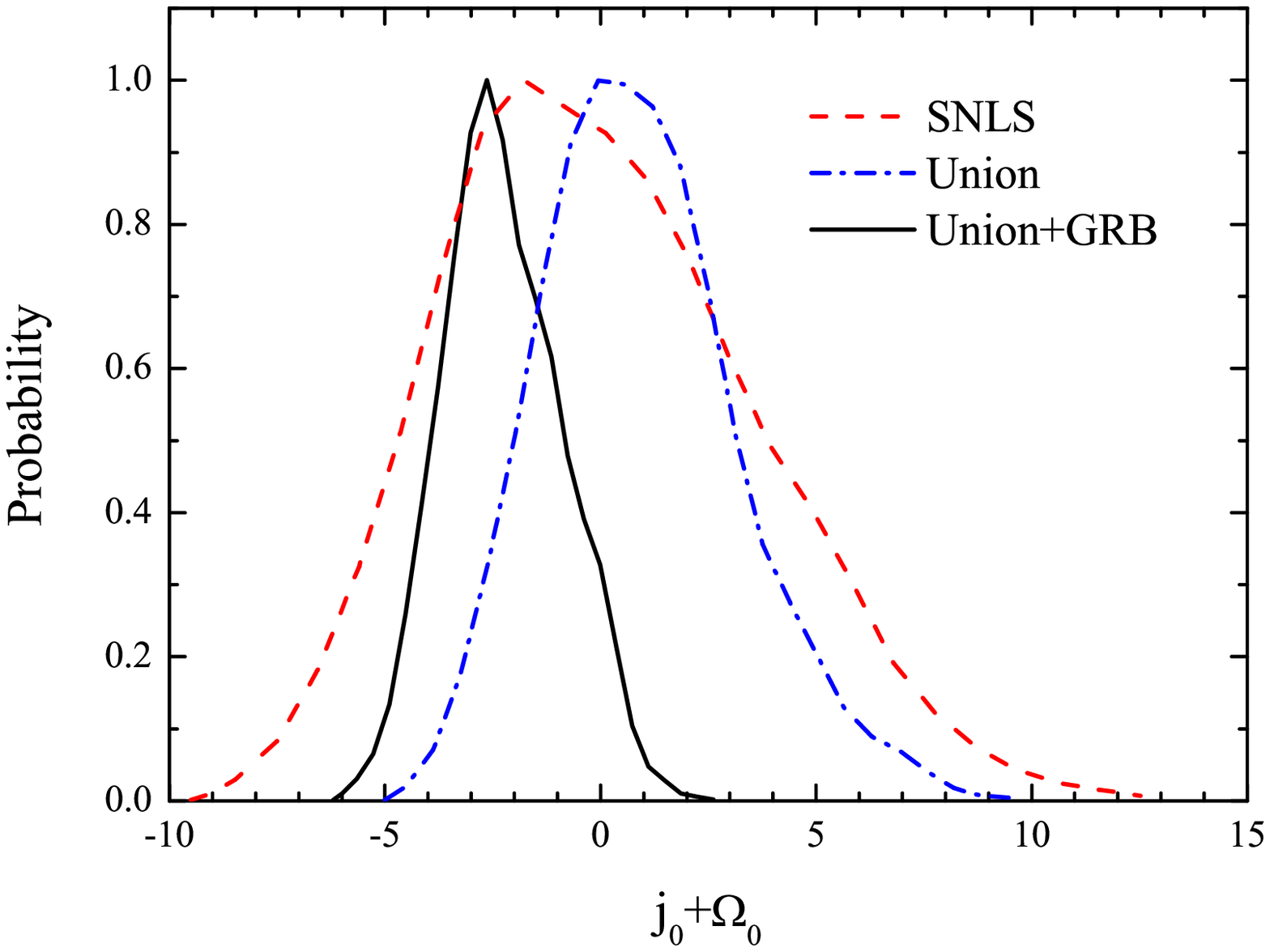}
\caption{One-dimensional likelihood distributions for the parameters
  $q_0$ and $j_0+\Omega_0$ from different data combinations in the
  Cosmography I case: SNLS (red dashed lines), Union (blue dash-dot
  lines) and Union+GRB (black solid lines).
  }
   \label{fig1}
\end{center}
\end{figure}

Firstly, we use the polynomial series of the logarithmic Hubble
relation up to second order, namely Cosmography I. In this case,
there are two free parameters: the deceleration $q_0$ and the jerk
parameter $j_0+\Omega_0$. In Fig.~\ref{fig1} we show the one
dimensional likelihood distributions for these parameters
as obtained from different data combinations.
Using the SNLS data only, we obtain the constraints on $q_0$ and
$j_0+\Omega_0$ at $68\%$ confidence level (1 $\sigma$):
\begin{equation}
q_0=-0.49\pm0.38~~,~~~j_0+\Omega_0=-0.52\pm3.52~,
\end{equation}
which is consistent with the previous work \cite{vc}. Furthermore,
the Union dataset gives more stringent $1\,\sigma$ constraints
on the parameters:
\begin{equation}
q_0=-0.58\pm0.24~~,~~~j_0+\Omega_0=0.91\pm2.21~,
\end{equation}
The error bars are shrunk by a factor of $\sim 1.5$, due to higher
accuracy of datasets and more samples at high redshifts. The value
of $q_0$ becomes smaller than that from SNLS data. Meanwhile,
$j_0+\Omega_0$ moves to a higher value, which is closer to the
expected value in the standard flat $\Lambda$CDM model, namely
$j_0+\Omega_0=2$ \cite{lucariello}.

Finally, we combine the Union compilation and GRB data together to
give the constraints in Cosmography I case. {We find that with this
data combination the mean value of $q_0$ will become larger, while
$j_0+\Omega_0$ goes to a smaller value than those of solely SNeIa
data.} The $1\,\sigma$ confidence levels are:
\begin{equation}
q_0=-0.24\pm0.19~~,~~~j_0+\Omega_0=-2.23\pm1.37~.
\end{equation}
The constraints have improved significantly.

For the reader's convenience we summarize in Table \ref{T1} the
constraints on these two parameters from different data combinations.

\begin{table}[h]
\begin{center}
\begin{tabular}{c|cc|cc}
  \hline
  \hline
  Data&\multicolumn{2}{c|}{SNLS}&\multicolumn{2}{c}{Union}\\
  Parameter &$q_0$&$j_0+\Omega_0$&$q_0$&$j_0+\Omega_0$\\
  \hline
  Best Fit&$-0.49$&$-0.36$&$-0.58$&$0.79$\\
  Mean&$-0.49\pm0.38$&$-0.52\pm3.52$&$-0.58\pm0.24$&$0.91\pm2.21$\\
  $\chi^2_{\rm
  min}$/d.o.f.&\multicolumn{2}{c|}{117.39/113}&\multicolumn{2}{c}{317.18/305}\\
  \hline
  \hline
  Data&\multicolumn{2}{c|}{GRB}&\multicolumn{2}{c}{Union+GRB}\\
  Parameter &$q_0$&$j_0+\Omega_0$&$q_0$&$j_0+\Omega_0$\\
  \hline
  Best Fit&$4.47$&$16.04$&$-0.23$&$-2.33$\\
  Mean&$4.02\pm2.05$&$18.31\pm24.86$&$-0.24\pm0.19$&$-2.23\pm1.37$\\
  $\chi^2_{\rm
  min}$/d.o.f.&\multicolumn{2}{c|}{74.74/67}&\multicolumn{2}{c}{402.65/374}\\
  \hline
  \hline
\end{tabular}
\end{center}
\caption{Constraints on the parameters of Cosmography I from
different data combinations ($1\sigma$ marginalized error bars).}
\label{T1}
\end{table}

It is interesting to stress that, for every set of data we considered for Cosmography I in Table \ref{T1}, it is not possible to attribute a definitive sign to the deceleration parameter $q_0$ up to a confidence level better than the $2\sigma$ one (a similar issue is also found in \cite{vc}).

\begin{figure}[htbp]
\begin{center}
\includegraphics[scale=0.4]{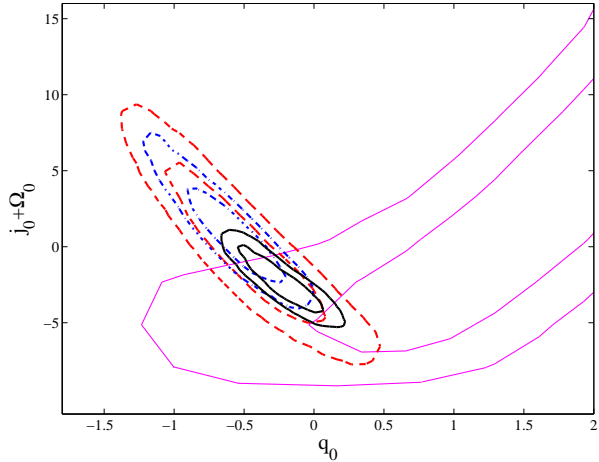}
\caption{Two-dimensional $1,\,2\,\sigma$ contours of parameters
($q_0$,$\,j_0+\Omega_0$) from different data combinations in the
Cosmography I case: SNLS (red dashed lines), Union (blue dash-dot
lines), Union+GRB (black solid lines) and GRB (magenta thick solid
lines). The GRBs now return a correlation in the
($q_0$,$\,j_0+\Omega_0$) plane \label{fig2}.}
\end{center}
\end{figure}

A further evidence emerging from the data fitting is associated to
the relation between $q_0$ and $j_0+\Omega_0$.  In fact, from
Fig.~\ref{fig2} it is evident that  while SNLS and Union give
consistent results (anti-correlated  $q_0$ and $j_0+\Omega_0$
parameters), GRB data seem to prefer a relatively strong
correlation between $q_0$ and $j_0+\Omega_0$ (albeit they give
rather weak constraints).

If confirmed this trend could be an indicator that high redshift data (i.e. those able to breakdown the degeneracy between different dark energy models) are showing a disagreement with the standard models as described by $\Lambda$CDM (for which one would instead expect an anti-correlation). However, an alternative explanation might be given by the fact that the early truncation of the cosmographic expansion characterizing our Cosmography I model is inadequate in properly fitting high redshift data of the GRB type. This feature is also suggested from the fact that the pure GRB data set seems to give rise to a high positive value of $q_0$. It is for this reason that we shall now deal with the next order truncation used in the Cosmography II model.

\subsection{Cosmography II}\label{subsec:C2}

With the accumulations of different observational data and the
improvements of the data quality, it is of great interest to
estimate the free parameters in the higher order polynomial terms.
In this paper we also try to estimate the snap parameter $s_0$ from
the current data. 

A first issue is related to the role of $\Omega_0$. While in the previous truncation (Cosmography I) it was always appearing in the expansion in combination with $j_0$ (so that $j_0+\Omega_0$ could be safely considered as a single parameter to be constrained), this is no more the case for the truncation Cosmography II (see Eq.~(\ref{eq:y-exp})). This is a potential problem because if one keeps $\Omega_0$ free then one does not have enough equations to determine all the independent parameters. We have then scanned the likelihood space when varying $\Omega_0$ and
found that for $0<\Omega_0<2$ one would basically obtain the same results. Henceforth, from now one we shall assume $\Omega_0=1$ in performing our analysis, since the effect of curvature can be safely neglected. 

\begin{figure}[htbp]
\begin{center}
\includegraphics[scale=0.36]{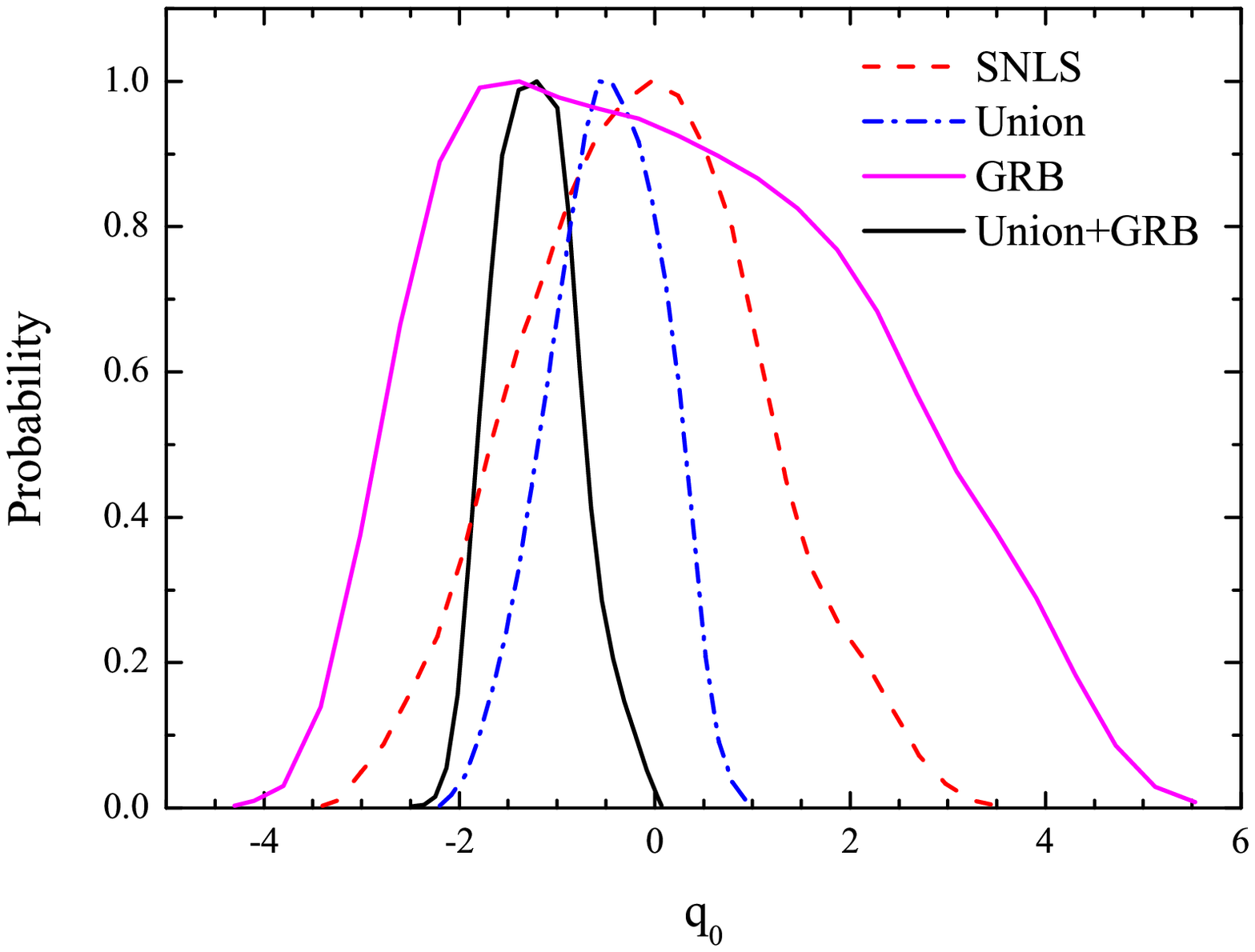}
\includegraphics[scale=0.36]{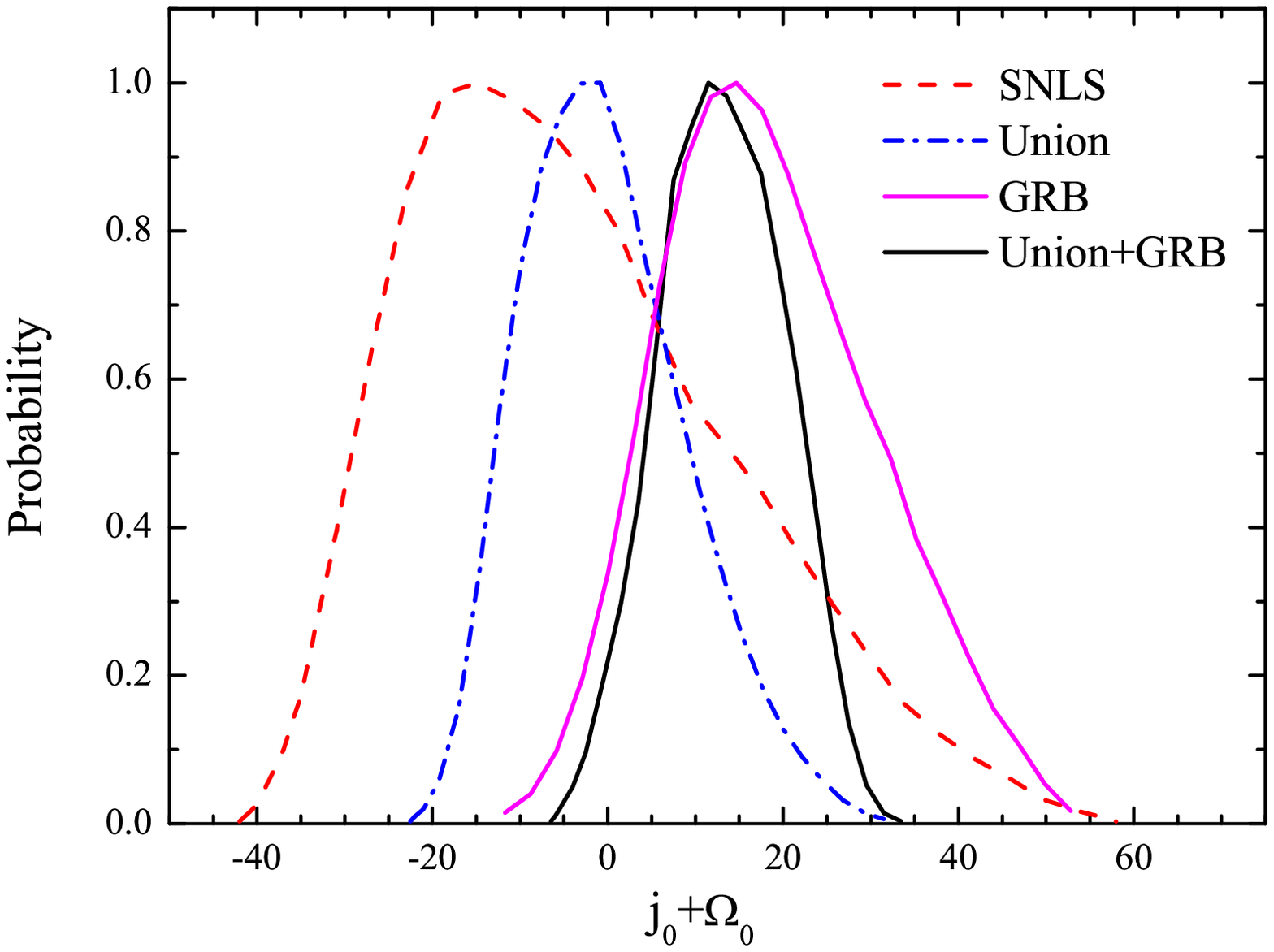}
\includegraphics[scale=0.36]{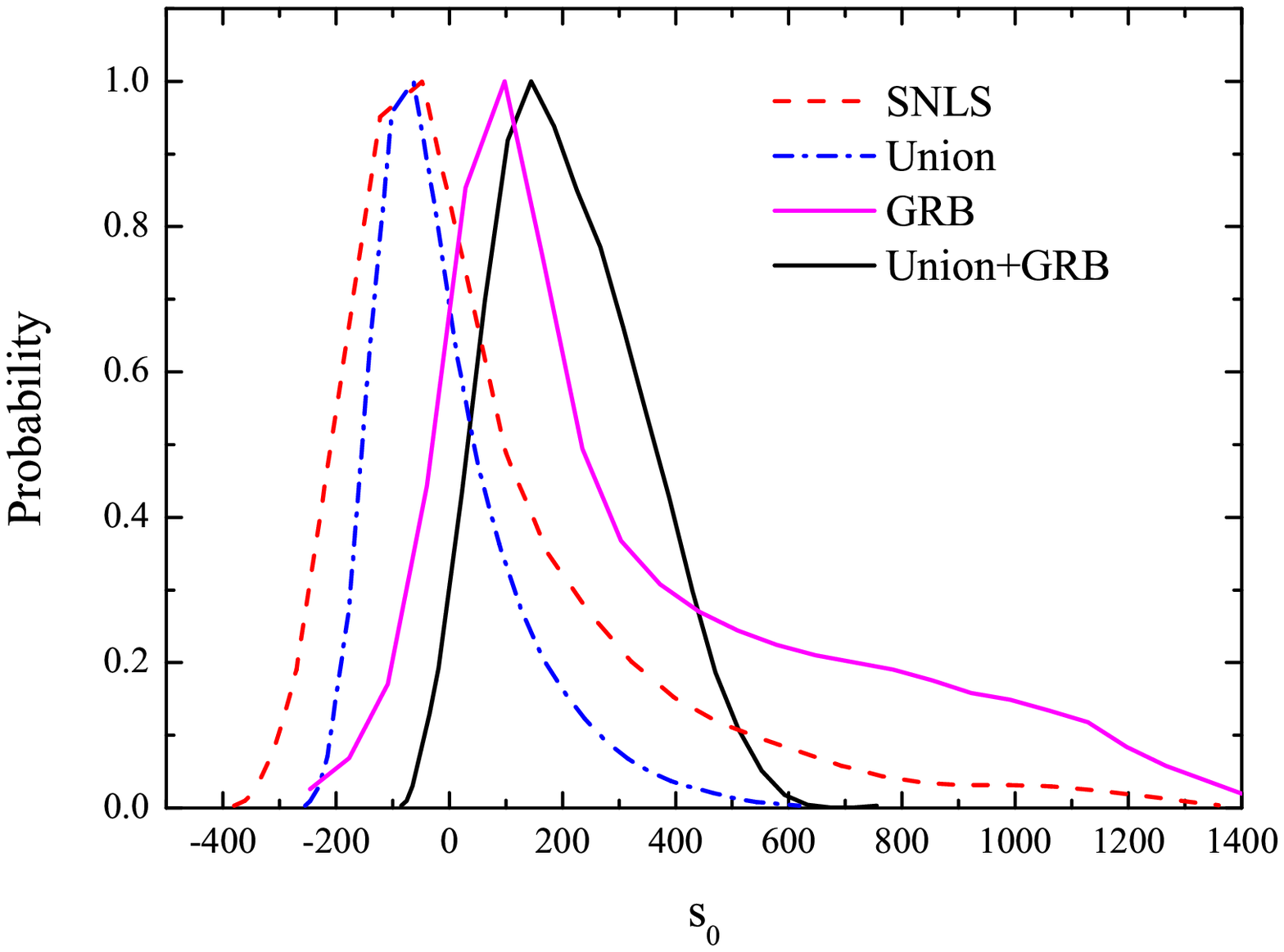}
\caption{One-dimensional likelihood distributions for the parameters
$q_0$, $j_0+\Omega_0$ and $s_0$ from different data combinations in
the Cosmography II case: SNLS (red dashed lines), Union (blue
dash-dot lines) and Union+GRB (black solid lines).}\label{fig3}
\end{center}
\end{figure}

In Fig.~\ref{fig3} we show the one dimensional likelihood
distributions for $q_0$ and $j_0+\Omega_0$ as obtained from
different data combinations.

{It is clear from the comparison of this figure with its analogue for Cosmography I, Fig.~\ref{fig1},  that in this case we do have a much more regular, and monotonic, improvement in the determination of the cosmological expansion parameters. In particular the use of the Union+GRB dataset does not lead anymore to reversal of the shift of the peak of the likelihood curves towards $O(1)$ positive values as it was happening with the model truncated at order $y^2$ and still gives the best constraints on the three parameters.
Furthermore we can now see from the third panel of Fig.~\ref{fig3} that it is only with the addition of the GRB data that the likelihood peak for the snap parameter $s_0$ noticeably changes. This is expected as only high redshift data can significantly constraint the higher order in the Hubble expansion (\ref{eq:y-exp}).}

In Table 2 we list the constraints on the parameters from different
data combinations when retaining terms up to third order in the polynomial expansion of the
logarithmic Hubble relation.
\begin{table}[h]
{\footnotesize
\begin{center}
\begin{tabular}{c|ccc|ccc}
  \hline
  \hline
  Data&\multicolumn{3}{c|}{SNLS}&\multicolumn{3}{c}{Union}\\
  Parameter&$q_0$&$j_0+\Omega_0$&$s_0$&$q_0$&$j_0+\Omega_0$&$s_0$\\
  \hline
  Best Fit&$-0.18$&$-6.42$&$-86.64$&$-0.36$&$-2.97$&$-61.87$\\
  Mean&$-0.12\pm1.15$&$-3.51\pm18.27$&$72.96\pm273.31$&$-0.50\pm0.55$&$-0.26\pm9.00$&$-4.13\pm129.79$\\
  $\chi^2_{\rm
  min}$/d.o.f.&\multicolumn{3}{c|}{117.32/112}&\multicolumn{3}{c}{317.07/304}\\
  \hline
  \hline
  Data&\multicolumn{3}{c|}{GRB}&\multicolumn{3}{c}{Union+GRB}\\
  Parameter&$q_0$&$j_0+\Omega_0$&$s_0$&$q_0$&$j_0+\Omega_0$&$s_0$\\
  \hline
  Best Fit&$-2.75$&$30.54$&$287.84$&$-1.26$&$13.64$&$211.27$\\
  Mean&$0.18\pm1.92$&$21.42\pm23.32$&$325.31\pm336.32$&$-1.22\pm0.41$&$13.22\pm6.85$&$214.51\pm127.84$\\
  $\chi^2_{\rm
  min}$/d.o.f.&\multicolumn{3}{c|}{72.37/66}&\multicolumn{3}{c}{395.24/373}\\
  \hline
  \hline
\end{tabular}
\end{center}}
\caption{Constraints on the parameters of Cosmography II from
different data combinations ($1\sigma$ marginalized error bars).}
\end{table}


In the case of Union data set improved by the adding of GRBs catalogue, we obtain for the first time a claim for a definitive negative $q_0$ within (even if marginally) $3 \sigma$.
Regarding this confidence level a comment is in order. 

There are several claims in the literature that confidence level $\geq 3\sigma$ on negative $q_0$ has been achieved with just Supernova data (see e.g.~\cite{Riess2007,ShapiroTurner}). Let us stress, however, that such claims are based on rather different parameterizations (mainly $z$-expansions of $q(z)$ or $w(z)$). In this sense our claim is limited to pure cosmographic expansion of the sort  (\ref{eq:y-exp}). Furthermore, it is important to keep into consideration that the use of $z$-based expansions with data in proximity and beyond $z=1$ tends to underestimate the errors with respect to the corresponding $y$-based expansion of the same parameterization \cite{vc_taylor}. For example, we have checked that using the same data and parametrization as in \cite{Riess2007} but expanding in the $y$ variable we can determine $q_0$ only up to $2\sigma$.

\begin{figure}[htbp]
\begin{center}
\includegraphics[scale=0.4]{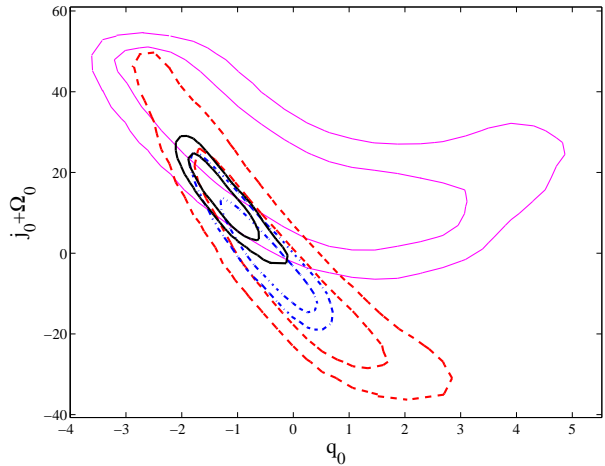}
\caption{Two-dimensional $1,\,2\,\sigma$ contours of parameters
($q_0$,$\,j_0+\Omega_0$) from different data combinations in the
Cosmography II case: SNLS (red dashed lines), Union (blue dash-dot
lines), Union+GRB (black solid lines) and GRB (magenta thick solid
lines). Now also using GRB date the anti-correlation is
maintained.\label{fig4}}
\end{center}
\end{figure}

Coming back to our analysis, it is also interesting to note that now (and in contraposition with Cosmography
I) GRB data also favors (as the Supernovae ones) an
anti-correlation between $q_0$ and $j_0+\Omega_0$, see
Fig.~\ref{fig4}. {This seems to strongly hint that the previously
found ($q_0$,$\,j_0+\Omega_0$) correlation in Cosmography I was an
artifact of the early truncation in the series (\ref{eq:y-exp}).}

\begin{figure}[htbp]
\begin{center}
\includegraphics[scale=0.325]{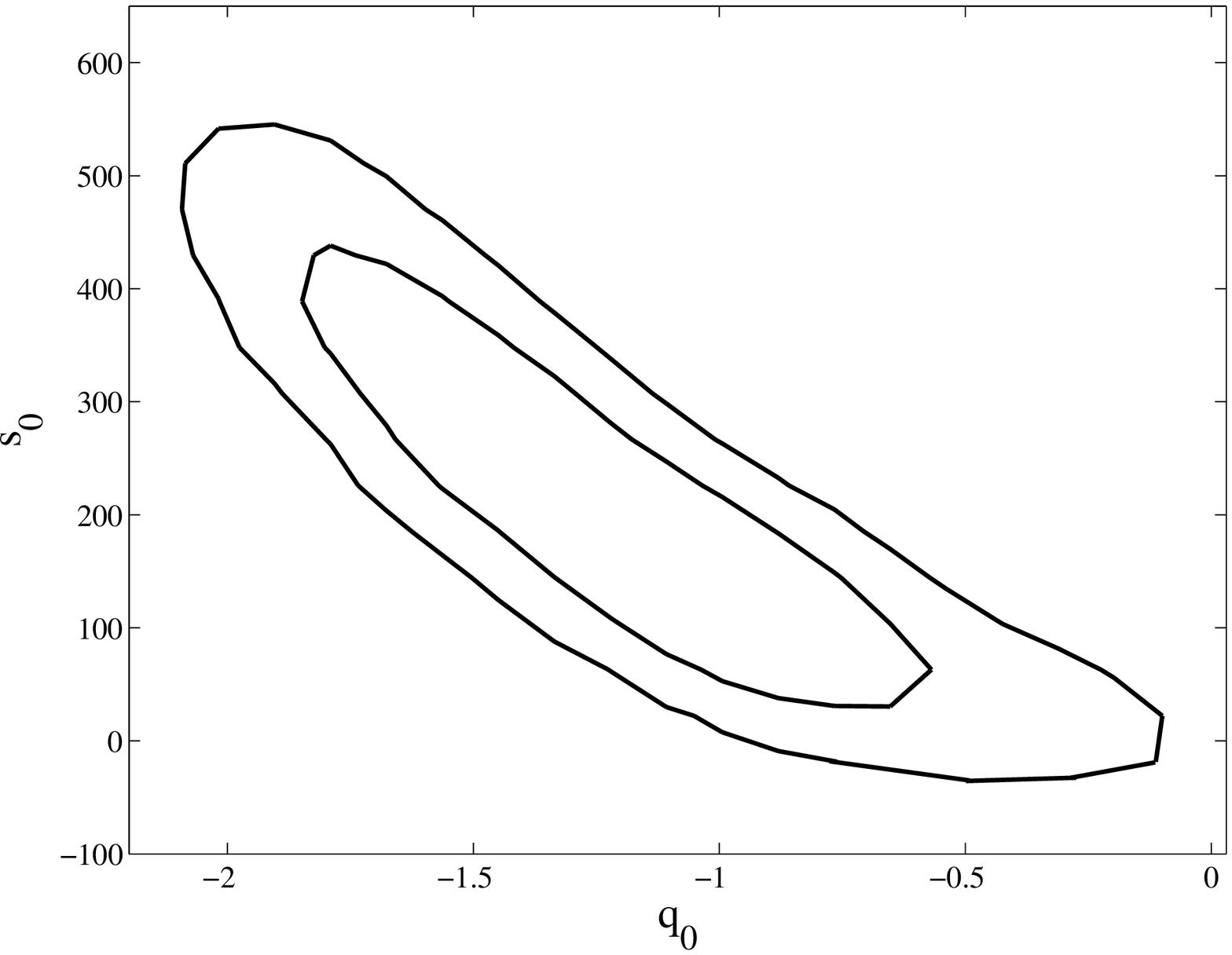}
\includegraphics[scale=0.325]{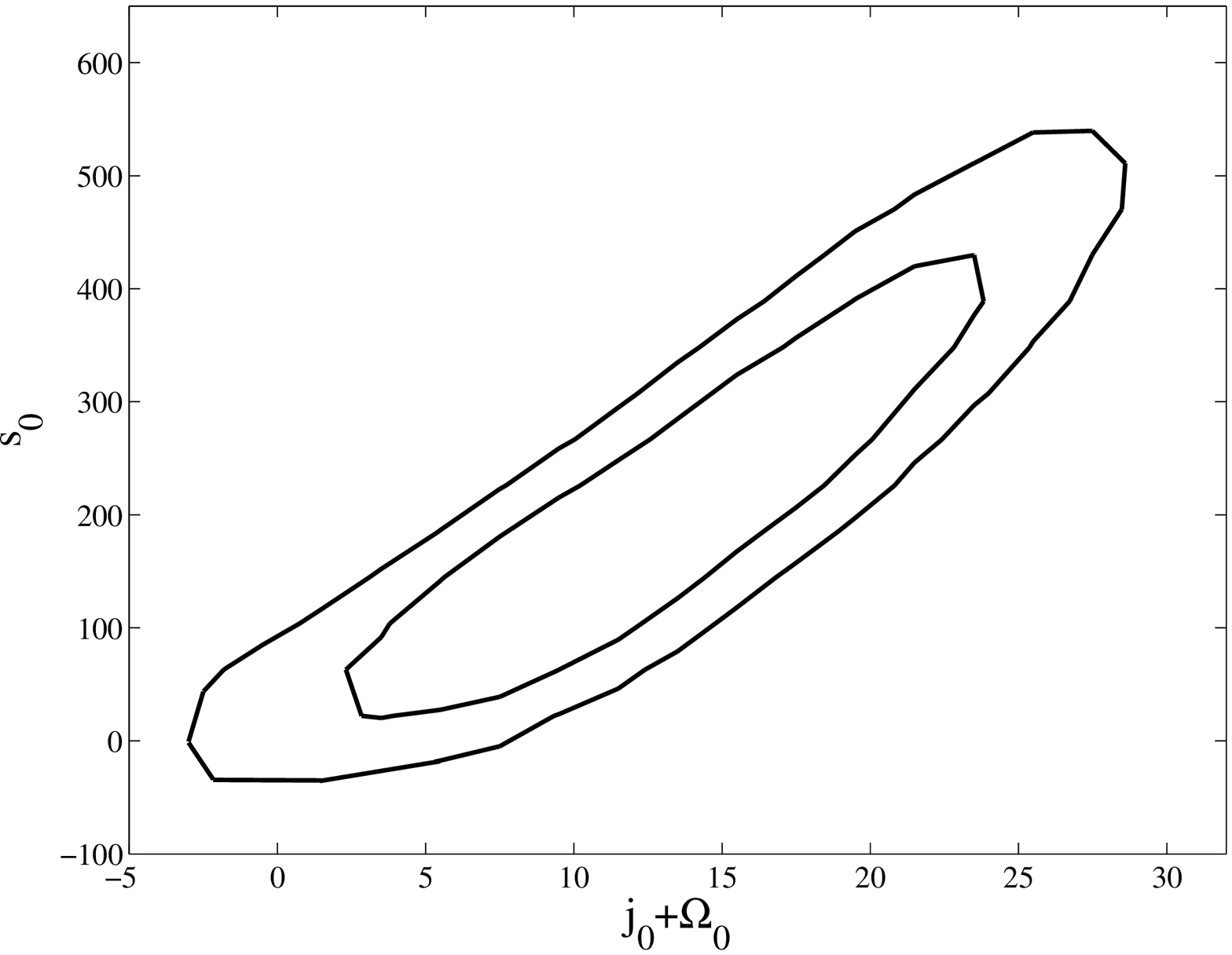}
\caption{Two-dimensional $1,\,2\,\sigma$ contours of parameters
($q_0$,$\,s_0$) and ($j_0+\Omega_0$,$\,s_0$) from Union+GRB
combination in the Cosmography II case.\label{fig5}}
\end{center}
\end{figure}

For completeness we also plot in Fig.~\ref{fig5} the two dimensional
contours between $q_0$, $j_0+\Omega_0$ and $s_0$. In this case
however the errors are still too large for a meaningful comparison
with theoretical cosmological solutions such as $\Lambda$CDM.

\subsection{Summary}

In section \ref{subsec:C1} and \ref{subsec:C2} we carried on the analysis of the Taylor
expansion of the luminosity distance in the $y$-redshift variable up
to second (Cosmography I) and third (Cosmography II) order. In
Ref.~\cite{vc} the authors correctly claimed that keeping the polynomial terms
up to the second order is a good approximation to the current SNLS data,
since the addition of any further term turns out to be of no
statistical relevance.

However, this work shows how a meaningful hope of estimating the
snap parameter $s_0$ becomes a concrete possibility with GRBs. Since
we are dealing with the truncation of a series, it is perfectly
reasonable that the higher is the redshift reached by data points,
the larger is the power of the polynomial we should consider to fit
them properly. Actually we have just see how an early truncation of
the series can lead to artifacts such as the
($q_0$,$\,j_0+\Omega_0$) correlation in Cosmography I when high
redshift data are included.

As a quantitative corroboration to this conclusion we can run, as
suggested in \cite{vc}, an $F$-test to establish which is
the last relevant term of the expansion we can reach. The $F$-test
provides a criterion of comparison between two nested models (in our
case two successive truncations of a Taylor series), identifying which
of two alternatives fits better the data. Supposing that the null
hypothesis implies the correctness of the first model, the test
verifies the probability to obtain that results fit the alternative
hypothesis as well. The less is this probability, the better is the
data fitting of the second model against the first
one. Quantitatively, one evaluates the $F$-ratio among the two
polynomials as \be
F=\frac{\l(\chi^2_1-\chi^2_2\r)/(n_2-n_1)}{\chi^2_2/(N-n_2)}\,, \ee
where $N$ is the number of data points, and $n_i$ represent the number
of parameters of the $i$-model. The $P$-value, \textit{i.e.} the area
subtended by the $F$-distribution curve delimited from the $F$-ratio
point, quantifies the viability of matching models as already
mentioned.

While the Cosmography II expansion, with both SNe and GRB,
is significantly better than Cosmography I (F-ratio=$6.99$, P-value$=.85\%$), we explicitly checked that
a truncation of the series to the fourth order is not favored, since
the decrease of $\chi^2$ value suggested by a possible Cosmography III is not
such to justify the introduction of a new parameter (F-ratio=$.08$, P-value$=77.12\%$).
In conclusion, Cosmography II seems to be currently the most
successful scheme for data fitting.

As a final remark let us note that using the Union+GRB dataset the
$\Lambda$CDM model returns a smaller $\chi^2$ than the two
Cosmographic expansions ($\chi^2_{\rm min}/d.o.f.=390.40/375$,
compared to Cosmography I which is $\chi^2_{\rm
min}/d.o.f.=402.65/374$ and Cosmography II which is $\chi^2_{\rm
min}/d.o.f.=395.24/373$). However, some caution is appropriate here
when interpreting $\chi^2$ values. First, $\Lambda$CDM is a solution
with one free parameter, $\Omega_m$, which is different from those
used in the cosmographic expansions (which, moreover, in the case of
Cosmography II are three). So the exact statistical significance of
the better $\chi^2$ of $\Lambda$CDM might be questionable. Second,
if indeed $\Lambda$CDM is the solution realized in nature then any
approximation to it based on a truncated Taylor series will provide
a higher $\chi^2$ (or at most a statistically indistinguishable) at any redshift in the presence of sufficiently accurate data. 
Nonetheless, the remarkably good
performance of $\Lambda$CDM at such high redshift, even with respect
to a Cosmographic expansion with more free parameters (which could
hence also reproduce the Hubble expansion of models with a
$z$-dependent equation of state for dark energy), could be taken as
a strong hint in favor of this specific solution.


\begin{figure}[htbp]
\begin{center}
\includegraphics[scale=0.36]{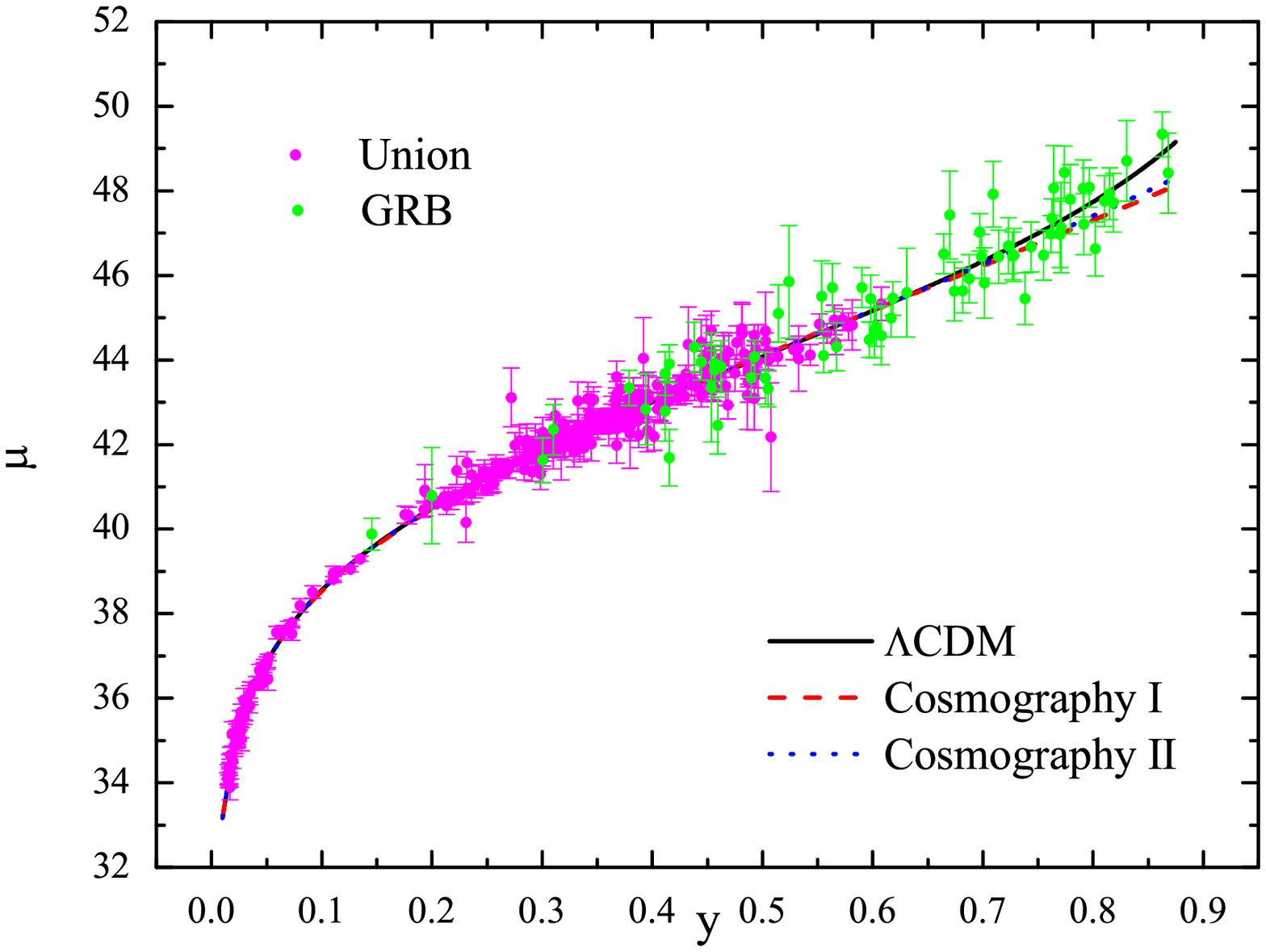}
\includegraphics[scale=0.36]{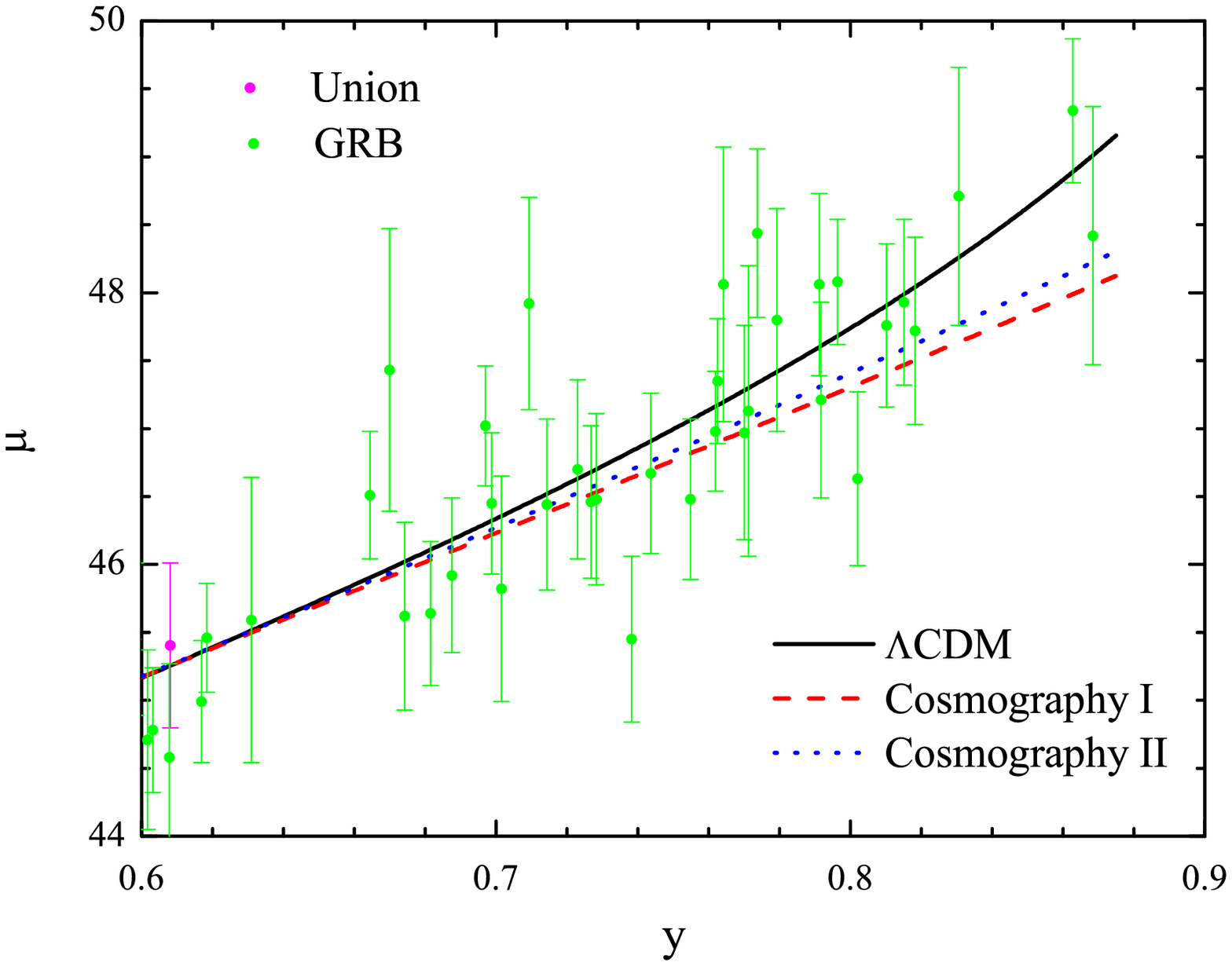}
\caption{Distance moduli for the best-fit values of three cases
  ($\Lambda$CDM, Cosmography I and Cosmography II), together with
  Union+GRB combination. In the left panel a zoom of the $0.6<y<0.9$ region is presented. The Union data set is shown in magenta and GRBs are in green.}
\end{center}
\end{figure}

\section{Future Perspectives}

We want now to present some forecasts for futuristic
but realistic mock data sets.  For this purpose we have performed
an analysis with respect to some fiducial model which is taken as the actual one realized in nature.
For concreteness we have taken such a model to be $\Lambda$CDM: $q_0=-0.55$,
$j_0=1$ and $s_0=-0.35$.

The projected satellite SNAP (Supernova/Acceleration Probe) would
be a space based telescope with a one square degree field of view
with $10^9$ pixels. It aims at increasing the discovery rate for SNeIa
to about $2000$ per year in the redshift range $0.2<z<1.7$. In this
paper we simulate about $2000$ SNeIa according to the forecast
distribution of the SNAP \cite{Kim:2003mq}. For the error, we follow
the Ref.~\cite{Kim:2003mq} which takes the magnitude dispersion
$0.15$ and the systematic error $\sigma_{\rm sys}=0.02\times z/1.7$.
The whole error for each data is given by:
\begin{equation}
\sigma_{\rm mag}(z_i)=\sqrt{\sigma^2_{\rm
sys}(z_i)+\frac{0.15^2}{n_i}}~,\label{snap}
\end{equation}
where $n_i$ is the number of Supernovae of the $i'$th redshift bin.
Furthermore, we add as an external data set a mock dataset of 400
GRBs, in the redshift range $0 < z < 6.4$ with an intrinsic
dispersion in the distance modulus of $\sigma_{\mu}=0.16$ and with a
redshift distribution very similar to that of Figure 1 of
Ref.~\cite{Hooper:2005xx}.


\begin{table}
\begin{center}
\begin{tabular}{c|ccc}
  \hline
  \hline
  Data&SN&GRB&SN+GRB\\
  \hline
  $q_0$&$0.42$&$1.11$&$0.33$\\
  $j_0+\Omega_0$&$7.97$&$11.23$&$4.92$\\
  $s_0$&$106.22$&$90.59$&$69.16$\\
  \hline
  \hline
\end{tabular}
\end{center}
\caption{The standard deviations, 1 $\sigma$ values, of the cosmographic parameters (uncertainties around the best fit value) expected from the
future measurements. }
\end{table}

In Table 3 we list the standard deviations of the free parameters in
two different cases. We can find the constraints on the parameters
from the future mock data with smaller error bars have improved by a
factor of $\sim2$, when comparison with the current constraints. When
including $s_0$, the constraints will be relaxed by a factor of $\sim
3$. Finally, the error of $s_0$ has been shrunk to $70$ by the future
SNeIa and GRB data. Possibly free parameters in the higher order terms
could be estimated by the future measurements, keeping in mind that
other potential probes can be used to extract meaningful distance
indicators such as the Alcock-Paczynski test using 21 cm at very high
redshift ($z>10$ e.g. \cite{xiaviel}), Baryonic Acoustic Oscillations
as measured by BOSS in SDSS-III (e.g \cite{mceisenstein}), the
Alcock-Paczynski for Lyman-$\alpha$ forest ($z=2-4$, \cite{mcdonald})
and measurements obtained with the shift from the Lyman-$\alpha$
absorption lines ($z=1.5-4.5$, \cite{Liske}).

\section{Conclusions}

Even though the intriguing era of ``precision cosmology" has now
started, the huge flow of data with increasing resolution is not yet
able to solve definitely the contest among different theoretical
proposals for the evolution of the universe at late times. The
suggestion carried on in this work is a return to a more conservative
approach to cosmology, in the meaning of a proper link between
observations and cosmographic parameters. In this sense cosmography
allows to regain a comprehensive bird's-eye view on the problem, since
it gives an unbiased interpretation to the collected data sets.

We have here included high redshift data ($z\gtrsim1$) to provide new
constraint on the main cosmographic parameters. The lack of validity
of the Taylor expansion of the luminosity distance in the usual
redshift definition has been circumvented, following the insight of
\cite{vc}, by the translation of the distance definition in a new
redshift variable $y$, spanning the range $(0, 1)$ whereas $z$ runs
into the interval $(0, +\infty)$; the improvement obtained with this
procedure is that now we are able to fit the data of any possible
observable candidate to be a cosmological ``standard candle", no
matter what its distance is.

Working with high redshift objects means taking into account both
distant SNe (up to $z\sim1.55$) and GRBs (up to $z\sim6.6$); in
particular, observations of GRBs are quickly approaching even larger
redshift (redshifts up to $z\sim10$ are expected in next few
years). This might make of GRBs the crucial players in the near
future, as such high redshifts will allow to determine the higher
order parameters of the cosmographic series with unprecedented
accuracy. Of course, we do not yet understand GRBs so well that we can
use them as proper standard candles. However, it has been proposed to
use correlations between GRB observables to standardize their
energetics \cite{grb1}. Furthermore, in this work we have also use the
insight of \cite{xiagrb} in order to solve the ``circular problem" in
the determination of the GRB redshifts.

We performed the calculation for two successive truncations of the
Taylor expansion of the luminosity distance (up to the third and
fourth order in $y$); further terms are, at the moment, not
statistically relevant, as shown by running an F-test.  The interest
about the possibility to reach the highest possible order is strictly
related to the main aim of cosmography: the ability to test and
discriminate meaningfully among competing cosmological models. Given
that most of the models on the market are build in order to recover
(or simulate) Dark Energy at low redshift, their expansion histories
are basically degenerate at late times. Breaking such a degeneracy
would imply knowledge of the early universe expansion curve and hence
would require, within a cosmographic approach, an accurate
determination of the higher order parameters jerk $j_0$ and snap
$s_0$. This can be achieved only via significant data at high
redshift.

Our results can be summarized in few points: we have found that the
inclusion of high $z$ GRB data might lead to misleading results for
early truncations of the cosmographic series as in Cosmography I.  We
showed that the cosmographic series truncated at the fourth order in
$y$, Cosmography II, is the most accurate and statistically
significant one given current data. The latter allows not only a more
accurate determination of the lowest order parameters $q_0$ and $j_0$
but also a first meaningful estimate of the snap parameter $s_0$. All
the results are compatible with $\Lambda$CDM within $2\sigma$. Most importantly, the complete set of Union and GRB data allows us to introduce for the first time a $q_0$ definitively negative in the limit of the $3 \sigma$ confidence level.

We have also discussed the future perspective to further ameliorate such
results with planned or proposed experiments. In this case we showed
that a further reduction of the error bars (by a factor two) will be
possible.

\acknowledgments {We would like to thank M.~Visser for interesting comments and useful suggestions. We also thank B.~Bassett and A.~Riess for useful criticism. VV wishes to thank warmly L.~Izzo for fruitful  discussions. MV is partly supported by ASI/AAE, INFN-PD51 and a PRIN
  by MIUR. Part of the numerical computations were performed using the
  Darwin Supercomputer of the University of Cambridge High Performance
  Computing Service (http://www.hpc.cam.ac.uk/), provided by Dell
  Inc. using Strategic Research Infrastructure Funding from the Higher
  Education Funding Council for England.}

\end{document}